\documentstyle[preprint,aps,eqsecnum,epsf,amsfonts]{revtex}
\begin{document}

{\tighten
\preprint{\vbox{\hbox{CALT-68-2009}
                \hbox{hep-ph/9507248}
		\hbox{\footnotesize DOE RESEARCH AND}
		\hbox{\footnotesize DEVELOPMENT REPORT} }}

\title{Leading logarithms of the $b$ quark mass \\[4pt]
  in inclusive $B\to X_s\,\gamma$ decay\footnote{%
Work supported in part by the U.S.\ Dept.\ of Energy under Grant no.\
DE-FG03-92-ER~40701.} }

\author{Anton Kapustin, Zoltan Ligeti and H.\ David Politzer}

\address{California Institute of Technology, Pasadena, CA 91125}

\maketitle

\begin{abstract}%
Part of the order $\alpha_s$ correction to the inclusive $B\to X_s\,\gamma$
photon spectrum is enhanced by $\log(m_b^2/m_s^2)$.  We discuss its origin and
sum the corrections proportional to $[\alpha_s\log(m_b^2/\mu^2)]^n$ to all
orders.  These are the calculable leading logarithms in the parton
fragmentation functions of a quark or gluon into a photon.  Although the gluon
fragmentation into a photon starts only at order $\alpha_s^2$, its contribution
is of the same order as the $s$ quark's in the leading log sum.  For not too
small values of the photon energy, the resummation yields a moderate
suppression.  In the standard model, the coefficient of the operator whose
matrix element gives rise to such terms is small.  A measurement of the photon
spectrum around $1\,$GeV would provide a theoretically clean determination of
$C_8$, the Wilson coefficient of the $b\to s\,g$ operator.
\end{abstract}

}%end tighten

\newpage

\section{Introduction}

Inclusive $B\to X_s\,\gamma$ decay is very sensitive to physics beyond the
standard model (SM) \cite{old,GSW,LLOb,NLO,BMMP,YY}.  As any flavor changing
neutral current process, it can only arise at one-loop level in the SM, and
therefore possible new physics can yield comparable contributions.  As the
recent CLEO measurement \cite{CLEO} excludes large deviations from the SM, a
next-to-leading order calculation is essential.  The absence of such a
calculation will soon become the major obstacle in improving the bounds on new
physics from this process.  Certain parts of this computation have already been
carried out \cite{NLO,AG}, but the technically most challenging parts are yet
to be calculated.

It was observed recently \cite{bsg1} that moments of the photon spectrum can be
predicted to order $\alpha_s$ accuracy by knowing the coefficients of the
effective Hamiltonian only to the presently available leading logarithmic
accuracy.  This will provide a model independent determination of the $b$ quark
pole mass (see also \cite{DSU}), that is, the $\bar\Lambda$ and $\lambda_1$
matrix elements of the heavy quark effective theory.

In this paper we point out that part of the next-to-leading order correction to
the photon spectrum induced by the $b\to s\,g$ operator is enhanced by
$\log(m_b^2/m_s^2)$.  We describe these contributions in section II.  In
section
III we resum the series of leading logarithms of the form
$[\alpha_s\log(m_b^2/\mu^2)]^n$.  We discuss the meaning of the infrared scale
$\mu$ and how nonperturbative phenomena enter the results.  The resummation of
leading logarithms does not eliminate the logarithmic dependence on the $b$
quark mass, which formally cancels the $\alpha_s$ suppression of this term.  In
section IV we discuss possible phenomenological implications.

\section{\lowercase{$\log(m_b^2/m_s^2)$} terms in the order
  \lowercase{$\alpha_{em}\,\alpha_s$} result}

In the standard model, $B\to X_s\,\gamma$ decay is mediated by penguin
diagrams.  QCD corrections to this process form a power series in the
parameter $\alpha_s\log(M_W^2/m_b^2)$, which is too large to provide a reliable
expansion.  Therefore, it is convenient to integrate out the virtual top quark
and $W$ boson effects (and possible new physics) at the $W$ scale, and sum up
the large logarithms using the operator product expansion and the
renormalization group.  We work with the operator basis and effective
Hamiltonian of Ref.~\cite{GSW}
\begin{equation}
H_{\rm eff} = -{4G_F\over\sqrt2}\, V_{ts}^*\,V_{tb}\,
  \sum_{i=1}^8 C_i(\mu)\, O_i(\mu) \,.
\end{equation}
For the present paper only two of these operators are directly relevant
\begin{eqnarray}\label{ops}
O_7 &=& {e\over16\pi^2}\, \bar s_\alpha\, \sigma^{\mu\nu}\, F_{\mu\nu}\,
  (m_bP_R+m_sP_L)\, b_\alpha \,, \nonumber\\
O_8 &=& {g\over16\pi^2}\, \bar s_\alpha\, \sigma^{\mu\nu}\, G_{\mu\nu}^a
  T_{\alpha\beta}^a\, (m_bP_R+m_sP_L)\, b_\beta \,,
\end{eqnarray}
where $\alpha$, $\beta$ are color indices, and
$P_{R,L}=\frac12(1\pm\gamma_5)$.  Throughout our discussion $C_7$ and $C_8$
will refer to the ``effective" Wilson coefficients \cite{BMMP} that include
the leading order contribution of the four quark operators.

We are interested in the order $\alpha_s$ correction proportional to $C_8^2$,
which is the order $\alpha_{em}$ corrections to $B\to X_s\,g$.
The finite part of the spectrum is given by
\begin{eqnarray}\label{spect88}
{{\rm d}\Gamma_{88}\over{\rm d}x}\bigg|_{x>0} = &&
  \Gamma_0\, {C_8^2\over9C_7^2}\, {\alpha_s\,C_F\over4\pi}\, \Bigg\{
  \bigg( {4+4r\over x-rx}-4+2x \bigg)\, \log{1-x+rx\over r} \\*
&& - {(1-x)\, [8 - (1-r)\,x\,(16-9x+7rx) + (1-r)^2\,x^3\,(1-2x)]\over
  x\,(1-x+rx)^2} \Bigg\} \,. \nonumber
\end{eqnarray}
Here $C_F=4/3$ in $SU(3)$,
\begin{equation}
\Gamma_0 = {G_F^2\, |V_{tb}V_{ts}^*|^2\, \alpha_{em}\, C_7^2\, m_b^5
  \over 32\,\pi^4}\, (1-r)^3\, (1+r) \,,
\end{equation}
is the leading order contribution to the $B\to X_s\,\gamma$ decay rate given
by the matrix element of $O_7$, and we introduced the dimensionless parameters
\begin{equation}
x = {2E_\gamma\over (1-r)\,m_b} \,,\qquad
r = {m_s^2\over m_b^2}\,.
\end{equation}
The variable $x$ corresponds to $E_\gamma/E_\gamma^{\rm max}$ in the free
quark decay model.
The extra factor of 9 suppression in eq.~(\ref{spect88}) results from the
square of the $b$ or $s$ quark's electric charge.

This contribution to the total $B\to X_s\,\gamma$ decay rate is infinite.  The
divergence is canceled by the virtual correction to the $B\to X_s\,g$ decay
mode (with no photon in the final state).  While soft photons could be summed
to all orders in $\alpha_{em}$ to make the $E_\gamma\to0$ limit smooth, in
finite orders of perturbation theory, it only makes sense to talk about the
total inclusive $B\to X_s\,\gamma$ decay rate with an explicit lower cut-off on
the photon energy.

The more interesting feature of the spectrum in eq.~(\ref{spect88}) is that it
is not finite in the $m_b/m_s\to\infty$ ($r\to0$) limit.  In particular, the
photon spectrum calculated in dimensional regularization with vanishing light
quark masses would contain a part proportional to $1/(D-4)$ for any value of
$E_\gamma$.  This is a consequence of the fact that a massless quark cannot be
distinguished from the same massless quark and one (or more) collinear massless
partons with the same total charge and energy.  The singularity is eliminated
once a nonzero quark mass is included, but it results in the above logarithmic
dependence of the spectrum on $r$.

Pictorially, this is simplest to understand by looking at the Dalitz plot in
the $r\to0$ limit (in the $E_\gamma - E_g$ variables), shown in Fig.~1.  The
shaded area represents the available phase space, the thick line denotes where
collinear singularities occur, and the black box indicates where the virtual
corrections contribute.  Fig.~1/a corresponds to the square of the invariant
matrix element of $O_7$.  In this case both the collinear singularity and the
virtual correction occur at $E_\gamma=m_b/2$.  For this contribution, there is
no problem whatsoever carrying out the calculation in the $m_s=0$ limit.  The
reason is that the kinematic variable $(1-x)$ regularizes all singularities,
and one finds the usual picture of Sudakov exponentiation near
$E_\gamma=m_b/2$.  However, the situation is more complicated for the
contribution of the square of the invariant matrix element of $O_8$, shown in
Fig.~1/b.  In this case, collinear singularities occur at any value of
$E_\gamma$; these are only canceled by the singular part of the virtual
correction in the total decay rate, but not in the photon spectrum.

If the hadronic final state $X_s$ formed a jet, by excluding photons within a
certain angular cut around the jet axis, the above discussed collinear
singularity would be eliminated.  However, at CLEO $B$ meson decay products are
distributed in angle.  Thus, when the experimental analysis excludes events in
which a photon and a charged particle overlap in the detector, this cut is not
related to the direction of the decay product $s$ quark at short distances.

\section{Resummation}

To identify the origin of the leading logarithms of the form
$[\alpha_s\log(m_b^2/\mu^2)]^n$, and to resum them to all orders, we observe
that the analytic structure and $r$-dependence of the $O_8$ contribution is
very similar to the photon structure function, as discussed in \cite{Witten}.
In the latter case one is interested in a hard photon scattering off another
(almost) on-shell photon, while in the present problem an on-shell photon is
created in the decay of a heavy particle.

We work to fixed (first) order in $\alpha_{em}$ and to all orders in
$\alpha_s$.  It is simplest to consider the discontinuity of the forward
scattering amplitude corresponding to intermediate states with an $s$ quark, a
photon, and any number of light quarks ($u$, $d$, and $s$) and gluons.  In an
axial gauge the leading logarithms come from those diagrams in which at order
$\alpha_s^n$ there are precisely $n$ ways of separating the external photon
legs from the other external legs by cutting two internal lines (see, {\it
e.g.}, \cite{factor}).  This implies that the result is given by the sum of two
sets of ladder diagrams.  One corresponds to the decay function of the $s$
quark into a photon, and the other to that of the gluon.  This second set of
ladder diagrams starts only at order $\alpha_s^2$, but also yields leading
logs.  From these arguments it also follows that only the square of the $O_8$
operator's invariant matrix element can yield these type of leading logs.

Decay functions are typically not calculable from first principles; moreover,
they are usually divergent in perturbation theory.  As the photon only couples
to massive partons (quarks), in a perturbative calculation the light quark
masses provide an infrared cut-off.  That is why eq.~(\ref{spect88}) in the
previous section contains a logarithmic dependence on $m_s$.  Higher order
corrections would also include terms proportional to $\log m_u$ and $\log m_d$.
However, in real QCD there is no such dependence on the light quark masses.
Non-perturbative decay functions are well-defined, measurable physical
quantities.

The crucial observation of Ref.~\cite{Witten} was that for a parton interacting
with an on-shell photon, the absolute normalization of the leading logarithms
can be calculated in perturbative QCD.  (In other cases only the $Q^2$
evolution is calculable.)  This large $Q^2$ (or large $m_b^2$) behavior is
determined by the matrix element of the photon operators between the external
photon states.  The contributions of quark and gluon operators are
logarithmically suppressed.  To sum the leading log ladders, one introduces an
infrared scale $\mu$, which is appropriately chosen to be a few times
$\Lambda_{\rm QCD}$.  The question of the $\mu$-dependence is then non-leading
in $\log m_b$.  Addressing it consistently involves a combination of
perturbative and non-perturbative ({\it i.e.}, decay functions deduced from
other experiments) terms of comparable magnitudes.  All strongly coupled long
distance effects are absorbed into the definitions of the decay functions.

The resummation of these leading logarithms is important, since $m_b\gg\mu$
implies $\log(m_b^2/\mu^2)\sim1/\alpha_s(m_b)$.  Only in this limit can
inclusive decay rates be calculated model-independently, based on an operator
product expansion \cite{CGG}.  The result of the resummation of the leading
logarithms is given by the sum of the decay functions of the $s$ quark and the
gluon into a photon
\begin{equation}\label{resm}
{{\rm d}\Gamma_{88}\over{\rm d}x}\bigg|_{\rm resummed} =
  \Gamma(b\to s\,g) \times [D^{s\to\gamma}(x)+D^{g\to\gamma}(x)]\,.
\end{equation}
Here $\Gamma(b\to s\,g)=\Gamma_0\,(\alpha_s\,C_F\,C_8^2/\alpha_{em}\,C_7^2)$
is the tree-level $b\to s\,g$ decay rate.
The calculable leading logarithmic part of the decay functions are
simplest to write down in terms of their moments ($n\geq2$) \cite{decayfn}
\begin{mathletters}\label{moments}
\begin{eqnarray}
\int_0^1 {\rm d}x\, x^{n-1}\, D^{s\to\gamma}(x) &=&
  {\alpha_{em}\log Q^2\over2\pi}\, V_n\,
  \bigg[ {e_s^2-\langle e_q^2\rangle\over 1+d_n^{qq}} +
  {\langle e_q^2\rangle\,(1+d_n^{gg})\over K_n} \bigg] \,, \\[4pt]
\int_0^1 {\rm d}x\, x^{n-1}\, D^{g\to\gamma}(x) &=&
  {\alpha_{em}\log Q^2\over2\pi}\, V_n\,
  {\langle e_q^2\rangle\, d_n^{qg}\over K_n} \,.
\end{eqnarray}
\end{mathletters}%
Here $V_n=\int_0^1{\rm d}x\,x^{n-2}\,[1+(1-x)^2]$, $e_s=-\frac13$ is the
$s$ quark's electric charge, and $\langle e_q^2\rangle$ denotes the average
of the squares of the light quarks' charges.  The $K_n$ and the $d_n$'s are
related to the anomalous dimension matrix of the quark and gluon operators
\cite{anomdim}.

In the absence of QCD,
$D^{s\to\gamma}(x)=(\alpha_{em}\,e_s^2/2\pi)\log Q^2\,[1+(1-x)^2]/x$ and
$D^{g\to\gamma}(x)=0$.  Substituting this into eq.~(\ref{resm}), we recover the
singular (in the $r\to0$ limit) part of the order $\alpha_{em}\,\alpha_s$
result discussed in section II.  The relevant part of eq.~(\ref{spect88}) is,
as promised,
\begin{equation}\label{sing88}
{{\rm d}\Gamma_{88}\over{\rm d}x}\bigg|_{\rm sing.} =
  \Gamma_0\, {C_8^2\over9C_7^2}\, {\alpha_s\,C_F\over2\pi}\,
  {1+(1-x)^2\over x}\, \log{1\over r}\,.
\end{equation}
It is not important that the $x\to1$ limit of eq.~(\ref{spect88}) is zero,
while that of eq.~(\ref{sing88}) is not.  The resummed photon spectrum vanishes
at $x=1$ anyway.  Away from $x=1$ the resummation of the leading logarithms
yields finite, non-zero corrections.  Although the gluon fragmentation into a
photon starts only at order $\alpha_s^2$, its contribution for not too large
photon energies is comparable to that of the $s$ quark.

The photon spectrum can be reconstructed from the moments in
eq.~(\ref{moments}) by inverse Mellin transformation.  We use three flavors in
the anomalous dimension matrix, as charm quarks are excluded from the final
state in the experimental analysis.  The sensitivity of the result to the
number of flavors in the $\beta$-function (three or four) is numerically very
small.  The $D^{g\to\gamma}$ term, being proportional to the square of the
quark charges, would be substantially bigger if charm were not excluded.  We
plot in Fig.~2.\ the ratio of the resummed $\Gamma_{88}$ contribution to the
photon spectrum divided by eq.~(\ref{sing88}).  For not too small values of the
photon energy the resummation suppresses the photon spectrum.  However, this
suppression is not very strong; it only exceeds a factor of two for about
$x>0.85$.  (The spectrum is enhanced for $x<0.3$.)

\section{Phenomenology and conclusions}

While we think the observations above are interesting even purely from a
theoretical point of view, we would like to discuss whether there are any
experimental implications.  Although enhanced by $\log(m_b^2/\mu^2)$, the
$\Gamma_{88}$ term discussed in this paper cannot become arbitrarily large as
compared to $\Gamma_0$.  The reason is that in the large $m_b$ limit the
running of $\alpha_s(m_b)$ cancels the logarithmic enhancement.  For a fixed
value of $m_b$ the $\mu\to0$ limit is not allowed.  On the one hand, it is cut
off by nonperturbative effects; on the other hand, the uncalculable parts of
the fragmentation functions can be absorbed in $\mu$, as long as $\mu$ is of
order $\Lambda_{\rm QCD}$.  Beyond perturbation theory, the $\log(1/r)$ term of
the order $\alpha_{em}\,\alpha_s$ result should be understood as
$\log(m_b^2/\mu^2)$, where $\mu$ is some scale of order $\Lambda_{\rm QCD}$.
Thus, it is also obvious that a measurement of the $\Gamma_{88}$ contribution
would not be sensitive to the strange quark mass.

For our numerical estimates we use $C_7(m_b)\simeq-0.31$ and
$C_8(m_b)\simeq-0.15$, corresponding to $m_t=175\,$GeV and
$\alpha_s(M_Z)=0.12$.  We also adopt $m_b=4.8\,$GeV and $0.3<\mu<1\,$GeV.  The
$\Gamma_{88}$ contribution to the energy region relevant for CLEO turns out to
be negligible.  The reason is mainly the overall $C_8^2/9C_7^2$ suppression.
On the other hand, any process that contributes to $B$ decays also contributes
to $B\to X\,\gamma$ via the fragmentation of any of the decay product partons
into a photon.  We estimate that the tree-level $W$-mediated decay $B\to\bar
u\,u\,\bar d\to X\,\gamma$ constitutes of order 1\% background to one of the
CLEO analyses that does not reconstruct a kaon.  This effect becomes more
significant and needs to be taken into account more precisely for any future
measurements of the $b\to d\,\gamma$ decay that aim at measuring $V_{td}$.

The $\Gamma_{88}$ contribution, however, dominates the photon spectrum below
about $1.2\,$GeV.  Measuring the inclusive $B\to X_s\,\gamma$ spectrum at such
low photon energies will probably remain impossible at CLEO.  The situation
may be better at asymmetric $B$ factories \cite{BaBar}.  But even if such
a measurement is possible in a bin around $1\,$GeV (say, $0.3<x<0.5$, that
corresponds to roughly $0.7<E_\gamma<1.2\,$GeV), the $B\to X_s\,\gamma$
branching fraction into this bin is about $6\times10^{-7}$.  Taking into
account expected experimental efficiencies for such an analysis, probably only
a handful of events can be detected (assuming $10^8$ $B$'s).  The reward for
this rather hard experimental analysis would be a theoretically clean
determination of $C_8$.

In conclusion, we pointed out that a part of the next-to-leading order result
for the inclusive $B\to X_s\,\gamma$ decay rate is proportional to
$\alpha_s\log(m_b^2/m_s^2)$.  Summing the leading logarithms of the form
$[\alpha_s\log(m_b^2/\mu^2)]^n$ to all orders in perturbation theory suppresses
(enhances) this contribution to the photon spectrum for large (small) values of
the photon energy.  In the large $m_b$ limit, formally, the logarithmic factor
cancels the $\alpha_s$ suppression of the $\Gamma_{88}$ contribution, as
compared to the leading order result $\Gamma_0$.  A measurement of the photon
spectrum about $E_\gamma\sim1\,$GeV would provide a theoretically clean
determination of $C_8$, the Wilson coefficient of the $b\to s\,g$ magnetic
moment type operator.

\acknowledgements
We thank Mark Wise for many useful discussions and Peter Cho and Ira Rothstein
for helpful conversations.  After this work was completed we received the
preprint \cite{AGnew}, which also discusses the $\Gamma_{88}$ contribution to
the photon spectrum.  Our result in eq.~(\ref{spect88}) agrees with eq.~(18) of
\cite{AGnew}.  Ref.~\cite{AGnew} mentions that soft photons could be summed to
eliminate the singularity near $E_\gamma=0$, but our main point is that
resumming collinear gluons is relevant for {\it any} photon energy.

{\tighten

\begin{figure}
\centerline{\epsfxsize=8truecm \epsfbox{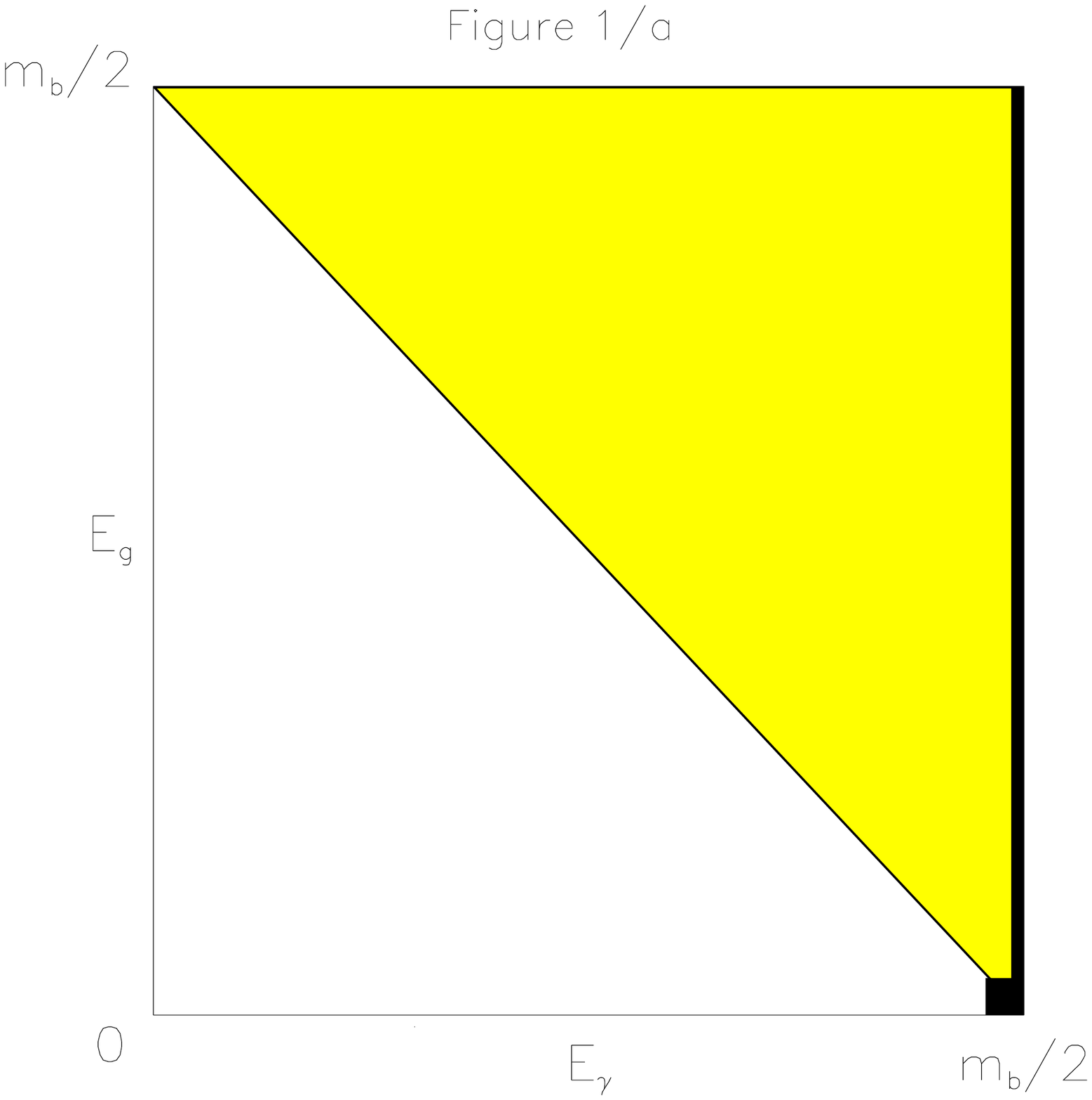}
  \epsfxsize=8truecm \epsfbox{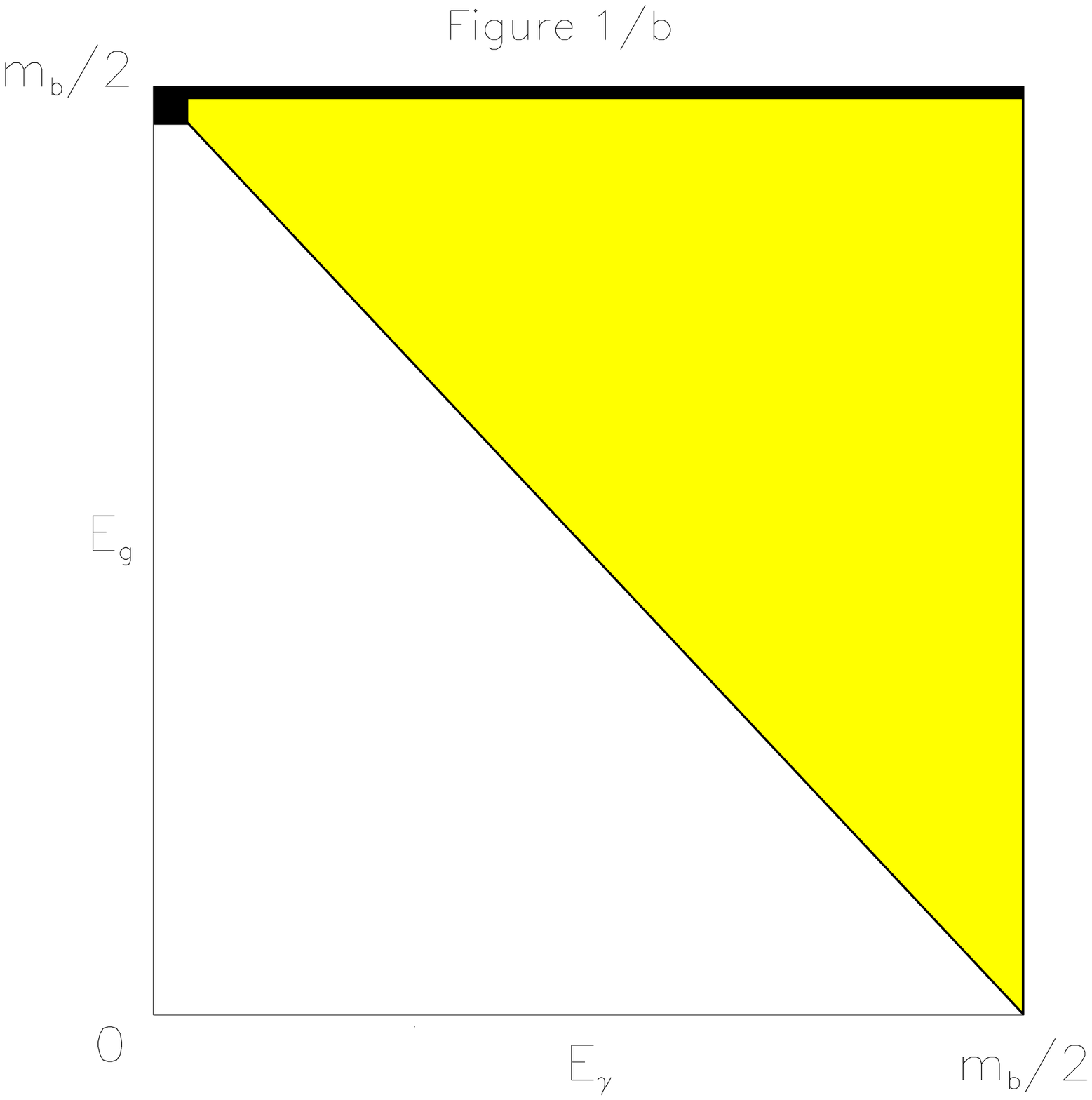}}
\caption[1]{Dalitz plot for the contribution of the square of the invariant
matrix element a) $O_7$, and b) $O_8$ to the decay $b\to s\,\gamma\,g$
in the $m_s\to0$ limit.
The shaded area is the available phase space, the thick lines denote
where collinear singularities occur, and the black box indicates
where the virtual corrections contribute.
}
\end{figure}

\begin{figure}
\centerline{\epsfxsize=16truecm \epsfbox{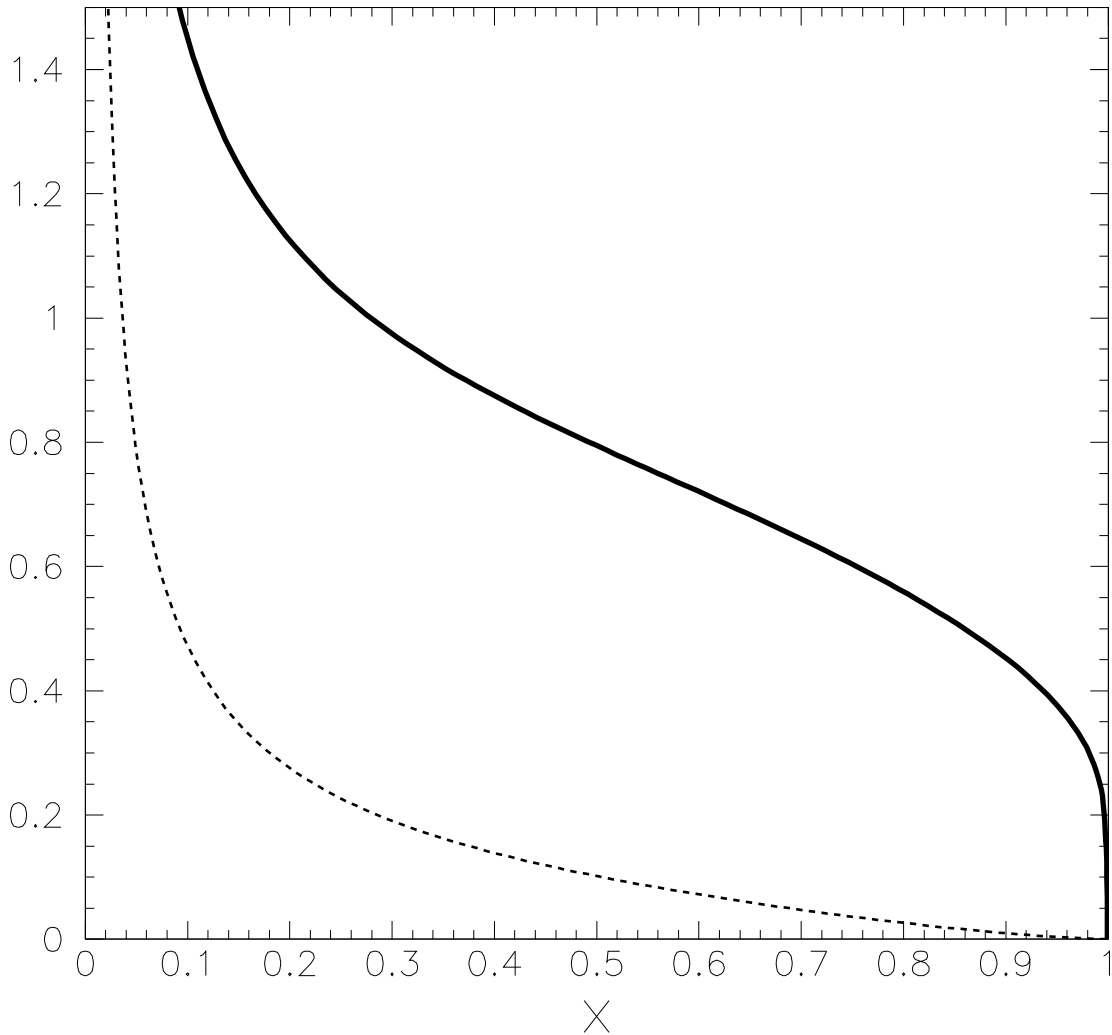} }
\caption[2]{The resummed $\Gamma_{88}$ contribution to the photon spectrum
divided by the leading log part of the Born term, eq.~(\ref{sing88}), as a
function of $x=2E_\gamma/(1-r)m_b$ (solid curve).  Dashed curve is the
gluon decay contribution, eq.~(3.2b).
}
\end{figure}

} %end tighten (references & figure captions)

\end{document}